\documentclass[onecolumn]{aastex6}

\shorttitle{Dynamical timescale inferred from chemical distribution in TMC-1}
\shortauthors{Choi et al.}

\begin{document}

\title{Dynamical timescale of precollapse evolution inferred from chemical distribution in the Taurus Molecular Cloud-1 (TMC-1) filament}

\author{Yunhee Choi and Jeong-Eun Lee}
\affiliation{School of Space Research, Kyung Hee University, 1732, Deogyeong-daero, Giheung-gu, Yongin-si, Gyeonggi-do 17104, Republic of Korea}

\author{Tyler L. Bourke}
\affiliation{Square Kilometre Array Organisation, Jodrell Bank Observatory, Lower Withington, Cheshire SK11 9DL, UK}

\and

\author{Neal J. Evans II}
\affiliation{Department of Astronomy, University of Texas at Austin, 2515 Speedway, Stop C1400, Austin, TX 78712-1205, USA \\ 
Korea Astronomy and Space Science Institute, 776 Daedeokdaero, Daejeon 305-348, Korea }

\email{jeongeun.lee@khu.ac.kr, yunhee.choi@khu.ac.kr}

\begin{abstract}
We present observations and analysis
 of the low-mass star-forming region, Taurus Molecular
Cloud-1 (TMC-1).
CS ($J$=2--1)/N$_2$H$^+$ ($J$=1--0) and C$^{17}$O ($J$=2--1)/C$^{18}$O ($J$=2--1) were observed with FCRAO (Five College Radio Astronomy Observatory) and SRAO (Seoul Radio Astronomy Observatory), respectively. 
In addition, {\it Spitzer} infrared data and 1.2~mm
continuum data observed with MAMBO (Max-Planck Millimetre Bolometer) are used.
We also perform chemical modeling to investigate the relative molecular
distributions of the TMC-1 filament.  Based on {\it Spitzer} observations,
there is no young stellar object along the TMC-1 filament, while five Class II
and one Class I young stellar objects are identified outside the filament.
The comparison between column densities calculated from dust continuum and
C$^{17}$O 2--1 line emission shows that CO is depleted much more
significantly in the ammonia peak than in the cyanopolyyne peak, while the
column densities calculated from the dust continuum are similar at the two
peaks. N$_2$H$^+$ is not depleted much in either peak. According to our
chemical calculation, the differential chemical distribution in the two
peaks can be explained by different timescales required to reach
the same density, i.e., by different dynamical processes.
\end{abstract}

\keywords{astrochemistry --- ISM: individual objects (TMC-1) --- ISM: molecules --- ISM: abundances}

\section{Introduction}
\label{sect-1}
Gravitationally bound starless cores, that is, prestellar cores, are cold
($T_{\rm k}\sim10$~K), dense ($n>10^4$~cm$^{-3}$) condensations of gas with
no sign of a central protostellar object, implying that they are the
earliest identifiable stage of star formation \citep[e.g.,][]{Shu87, Lee99,
Crapsi05, Ward-Thompson07}.  Therefore, prestellar cores are the places
that we can study the initial conditions of star formation such as the
distribution of density, temperature, and velocity \citep{Benson89,
Ward-Thompson94}.  Prestellar cores have a constant density in the inner
part and a decreasing density at larger radii \citep{Ward-Thompson94,
Ward-Thompson99, Ward-Thompson00, Andre96}, which can be well described
with a Bonnor-Ebert sphere density profile \citep{Evans01}.  Dust grain
temperatures decrease toward the center of prestellar cores \citep{Evans01,
Zucconi01, Ward-Thompson02, Crapsi07, Harju08, Nielbock12} since the interstellar radiative field (ISRF) is the only heating sources of prestellar cores.

Although there are several model scenarios to explain core formation, they
can be divided into two extreme paradigms: slow quasistatic contraction
and dynamic turbulent processes.  In the former process, a core
gradually becomes condensed toward the center through ambipolar diffusion,
increasing the ratio of core mass to magnetic flux. 
As a core switches from magnetically subcritical to a
supercritical condition, the core can collapse
gravitationally \citep{Mouschovias91, Mouschovias99, Shu87}.  
If compressible turbulence dominates in a prestellar core, the
turbulence will decline in a short time \citep[e.g.,][]{MacLow98} and
collapse will happen. This means that the prestellar core disappears
quickly or rapidly collapses. The detailed modeling results of this process
suggest that the cores live for only one or two free-fall times
\citep[e.g.,][]{Ballesteros-Paredes03, Vazquez-Semadeni05}.
Demographic studies indicate lifetimes that decline with 
increasing density from near 10 free-fall times at $10^3$ cm$^{-3}$ 
to a few free-fall times, or about 0.5 Myr, once the mean
density rises above $2 \times 10^4$~cm$^{-3}$, and one free-fall time
once the mean density is above $10^6$~cm$^{-3}$
\citep{Ward-Thompson07, Enoch08, Andre14}.

Studies show that the chemical compositions of dark cloud cores can
be a useful tracer of their evolutionary status in a given dynamical
process, as the chemistry is time dependent.  
The dynamical evolutionary stage
can be inferred by the density structure of a core, which can be obtained
through the dust continuum emission.  However, one of the important
properties of dynamical processes is the velocity distribution, which is
traced only by line profiles within the core.  Molecular line profiles are
affected by not only density and temperature profiles but also the
abundance structure in the core.  Therefore, it is necessary to understand
the chemical distribution within cores \citep[e.g.,][]{Mizuno90, Zhou93,
Choi95, Bergin95, Gregersen97, Gregersen00, Mardones97, Aikawa01, Lee03,
Lee04, DiFrancesco01, Belloche02}.

Taurus Molecular Cloud-1 (TMC-1) is an interesting source for exploring the
chemistry of dark clouds.  It is a cold and dense cloud with a narrow
filamentary structure (5$\arcmin\times15\arcmin$) in the Taurus dark-cloud
complex at a distance of $\sim$140~pc 
\citep{Elias78, Kenyon94, Torres12} extending
from the southeast to northwest direction, with an infrared source (IRAS
04381+2540) located outside the northwest part.  The TMC-1 filament has
been observed in many molecular lines. 
According to those molecular observations, carbon-chain molecules such as CCS, C$_4$H, and HC$_3$N are abundant in the southeast part (cyanopolyyne
peak) of the TMC-1 filament, and NH$_3$ are N$_2$H$^+$ are abundant in
its northwest part (ammonia peak). This peculiar chemical feature of TMC-1
has been an interesting problem, so there have been many studies attempting
to understand the chemical distributions in TMC-1
\citep[e.g.,][]{Churchwell78, Little79, Wootten80, Toelle81, Guelin82,
Snell82, Schloerb83, Olano88, Hirahara92, Suzuki92, Pratap97, Hirota98,  Markwick00, Markwick01, Saito02, Hirota03,
Suutarinen11}.

Many studies have concluded that the
cyanopolyyne peak is at an earlier evolutionary stage than the ammonia
peak, and thus the cyanopolyyne
peak (southeast part) is less chemically evolved than the northwest part
based on the chemical models and molecular line observations such as CCS,
HC$_5$N, C$^{34}$S, NH$_3$, C$^{34}$S, DCO$^+$, H$^{13}$CO$^+$, and CH.
The estimated chemical evolutionary age
difference between the two peaks is more than 10$^5$~yr
\citep[e.g.,][]{Suzuki92, Hirahara92, Hanawa94, Howe96, Pratap97, Saito02,
Suutarinen11}.

\citet{Hirahara92} showed that the H$_2$ density of the southeast part was
lower than that of northwest part by a factor of 10 using C$^{34}$S. They
suggested that the cyanopolyyne peak was in an earlier stage of chemical
evolution than the ammonia peak because of the higher density at the
ammonia peak. 
\citet{Pratap97} observed several molecular species and
suggested two possibilities for the chemical difference along the TMC-1
filament. One was a small variation in the C/O ratio from 0.4 to 0.5.
Another was that differences in density,
which varies by a factor of 2 along
the filament, could explain the difference of chemical distribution in
the TMC-1 filament.  The latter was consistent with the result by
\citet{Hirahara92} because the higher density could drive chemical
processes more rapidly.  Recently, \citet{Suutarinen11} presented a CH
abundance gradient in TMC-1, from $\sim$1.0$\times$10$^{-8}$ in the
northwestern part to $\sim$2.0$\times$10$^{-8}$ in the southeastern part of
TMC-1, and they noted that 
the CH column density peaks close to the cyanopolyyne peak.  They
suggested that the southeastern part in the vicinity of the cyanopolyyne
peak had an extensive low-density envelope based on the comparison of the
dust continuum maps from SCUBA 850 and 450~$\micron$ \citep{Nutter08} and
the modelling of molecular lines \citep{Hirahara92, Pratap97}, implying
that the cyanopolyyne peak was in the early stage of dynamical evolution and
this peak was also chemically less evolved than the ammonia peak because of
the lower density.

\citet{Markwick00, Markwick01} proposed another explanation for
molecular abundance differences along the TMC-1 filament. 
They investigated the possibility that
the gradients were produced by spontaneous explosive desorption of UV-photolyzed ice mantles,
triggered by the heating of dust grains and their ices, through grain-grain collisions
induced by MHD waves from IRAS 04381+2540 near the ammonia peak.
The chemical processes occurred first in the ammonia peak and later in the cyanopolyyne peak,
which induced the gradient of molecular abundances along the TMC-1 filament.
According to their chemical models, the ammonia peak is 1.5$\times$10$^5$~yr older than the cyanopolyyne peak, 
assuming the speed of the MHD wave of 2~km~s$^{-1}$.

A very different explanation for the chemical distribution along the TMC-1
filament might be possible. 
According to gas-phase chemical models, carbon-chain species such as CCS form before atomic carbon mainly reacts with atomic oxygen, which drives all carbon-bearing species into CO.
When the depletion of CO (and even N$_2$H$^+$) is very significant, abundances of carbon-chain species increase again, which is called a late-time secondary abundance peak of carbon-chain species \citep{Ruffle97, Ruffle99, Li02, Lee03}.
In this case, the cyanopolyyne peak might be chemically older than the ammonia peak,
unlike the conclusion from the rest of the studies.  Therefore, we need to
calculate accurate depletion factors of N$_2$H$^+$ and CO in the two peaks
using accurate H$_2$ column densities. 

In this study, we present new data from
the FCRAO (Five College Radio Astronomy Observatory)
14-m Telescope and the SRAO (Seoul Radio Astronomy Observatory) 6-m
Telescope, along with data from the {\it Spitzer} Space Telescope 
and MAMBO (Max-Planck Millimetre Bolometer) at the IRAM 30-m Telescope. 
We combine these data sets
to study the reason for the chemical differentiation along the filament.
We determine the degree of molecular depletion by comparing column
densities calculated from dust continuum emission and molecular line
emission, since the dust continuum at (sub)millimeter traces the total
material along the line of sight very well.  We also carry out chemical
calculations to study the abundance evolution of CO and N$_2$H$^+$ 
with different assumptions about the evolution of the core densities.

Observations of TMC-1 are summarized in Section~\ref{sect-2}. In
Section~\ref{sect-3} we show the observational results and simple analysis.
Section~\ref{sect-4} presents the column densities using several molecules
and dust continuum emission. Section~\ref{sect-5} describes the chemical
modeling sequence and results. Finally, we discuss our results in
Section~\ref{sect-6}.

\section{Observations}
\label{sect-2}

\subsection{Five College Radio Astronomy Observatory (FCRAO) Observations}
Maps of CS ($J$=2--1) and N$_2$H$^+$ ($J$=1--0) toward TMC-1 were
made in 2003 January with the 14-m telescope of the Five College Radio Astronomy Observatory
(FCRAO).  The 32-element heterodyne receiver SEQUOIA was used with the auto-correlator configured
with a band width of 25 MHz over 1024 channels.
The total mapping times were  237 hours  and 130 hours for CS ($J$=2--1)  and N$_2$H$^+$ ($J$=1--0), respectively.
The reference center of maps was ($\alpha$, $\delta$)=(4$\rm^h$41$\rm^m$44.0$\rm^s$, +25$\arcdeg$42$\arcmin$22.0$\arcsec$).

\subsection{Seoul Radio Astronomy Observatory (SRAO) Observations}

We observed TMC-1 in C$^{17}$O ($J$=2--1) and C$^{18}$O ($J$=2--1) with the
6-m telescope of Seoul Radio Astronomy Observatory (SRAO) in 2009 February.
The autocorrelator was configured with a band width of 100~MHz over 2048
channels. The pointing uncertainty was less than 10$\arcsec$ in both
azimuth and elevation. 
We observed the positions of the cyanopolyyne peak and the ammonia peak in the TMC-1 filament.
The total integration time for each position was 30 minutes.

Table~\ref{tbl-1} summarizes basic observational parameters of the FCRAO and SRAO data,
including the frequency ($\nu$), velocity resolution ($\delta v$),
the FWHM beam size ($\theta_{\rm mb}$), the main beam efficiency
($\eta_{\rm mb}$), rms noise level at the given velocity resolution, and the observing date.

\begin{deluxetable*}{lcccccccc}
\tablecaption{List of Observed Lines.\label{tbl-1}}
\tablecolumns{9}
\tablewidth{0pt}
\tablehead{
\colhead{Molecule} & \colhead{Transition} & \colhead{Telescope} & \colhead{$\nu$} & \colhead{$\delta v$} & \colhead{$\theta_{\rm mb}$} & \colhead{$\eta_{\rm mb}$} &  \colhead{rms} & \colhead{Observing Date} \\
\colhead{} & \colhead{} & \colhead{} & \colhead{(GHz)} & \colhead{(km s$^{-1}$)} & \colhead{(arcsec)} & \colhead{} &\colhead{(K)} & \colhead{}}
\startdata
N$_2$H$^+$  & $J$=1--0 & FCRAO 14~m  & 93.174    & 0.08   & 58 & 0.5 & 0.1 &2003 Jan. \\
CS                  & $J$=2--1 & FCRAO 14~m & 97.981    & 0.07   & 55 & 0.5 & 0.1 & 2003 Jan. \\
C$^{18}$O     & $J$=2--1 & SRAO 6~m      & 219.560   & 0.13 & 48 & 0.57 & 0.04 & 2009 Feb. \\
C$^{17}$O     & $J$=2--1 & SRAO 6~m      & 224.714   & 0.13 & 48 & 0.57 & 0.04 & 2009 Feb. \\
\enddata
\end{deluxetable*}

\subsection{The {\it Spitzer} Space Telescope Observations}

The {\it Spitzer} Legacy Program ``From Molecular Cores to Planet Forming
Disks'' \citep[c2d;][]{Evans03} observed TMC-1 at 3.6, 4.5, 5.8, and
8.0~$\micron$ with the Infrared Array Camera \citep[IRAC;][]{Fazio04} on
2004 October 8 and at 24 and 70~$\micron$ with the Multiband Imaging Photometer
for {\it Spitzer} \citep[MIPS;][]{Rieke04} on 2004 September 25.  The pixel
sizes are 1$\arcsec$.2 for all IRAC bands and 2$\arcsec$.45 at MIPS
24~$\micron$. These images were centered at ($\alpha$, $\delta$)=
(4$\rm^h$41$\rm^m$14$\rm^s$, +25$\arcdeg$59$\arcmin$31$\arcsec$) for the
3.6 and 5.8 $\micron$ bands and (4$\rm^h$41$\rm^m$19$\rm^s$,
+25$\arcdeg$52$\arcmin$49$\arcsec$) for the 4.5 and 8.0 $\micron$ bands.
MIPS image was centered at ($\alpha$,
$\delta$)=(4$\rm^h$41$\rm^m$21$\rm^s$, +25$\arcdeg$55$\arcmin$46$\arcsec$).
For detailed descriptions of data reduction refer to \citet{Evans07}.

\subsection{Max-Planck Millimetre Bolometer (MAMBO) Observations}
The observations of the 1.2~mm thermal dust continuum emission toward TMC-1
were performed with the IRAM 30-m-telescope on Pico Veleta (Spain) using
the 117-receiver MAMBO-2 (240$\arcsec$ diameter) camera \citep{Kreysa99}.
The beam size on the sky was 11$\arcsec$, and the effective frequency is
250~GHz with half sensitivity limits at 210~GHz and 290~GHz. TMC-1 was
observed between the summer of 2002 and the winter of 2003/2004.  The
detailed process of the data reduction of MAMBO observations was explained in
\citet{Kauffmann08}, where these data were first presented.

\section{Observational Results \& Basic Analysis}
\label{sect-3}
We present the relevant data sets in this section, starting with the
infrared and submillimeter continuum data, which establish the
environmental context for the molecular line observations.

\subsection{Infrared Observations}

\begin{figure}
\epsscale{0.8}
\plotone{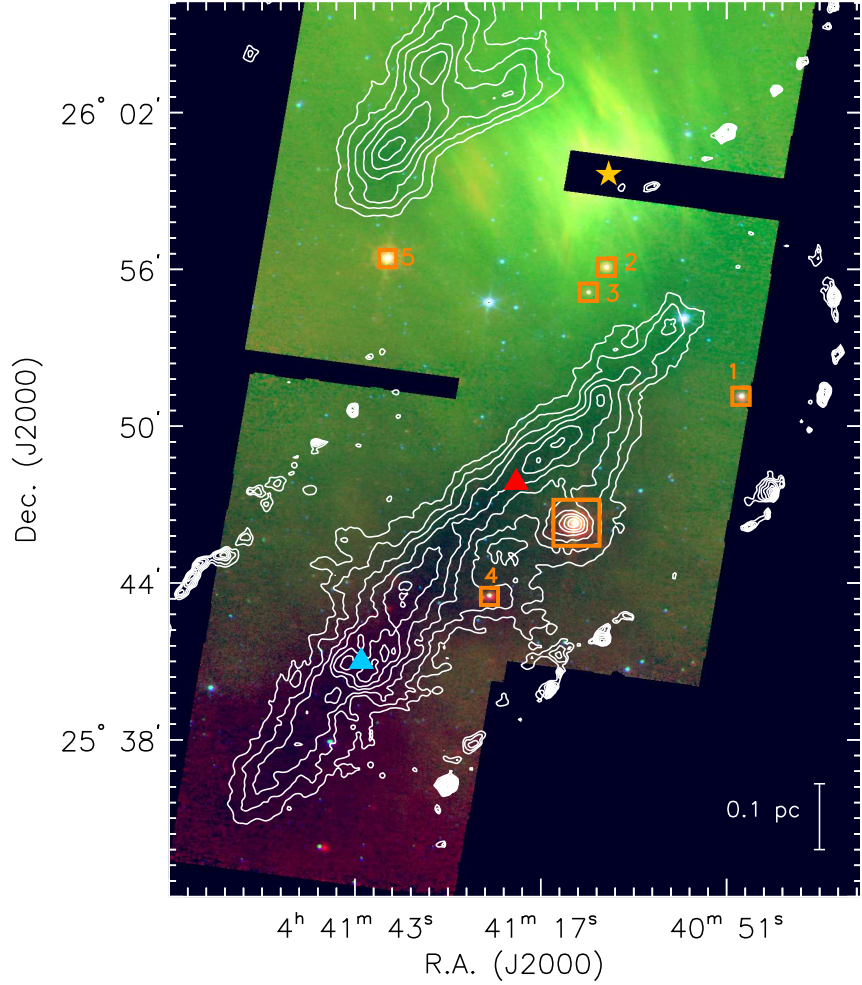}
\caption{The MAMBO 1.2~mm dust continuum emission (contours) on top of
three-color composite {\it Spitzer} image of TMC-1. The IRAC 4.5~$\micron$,
8.0~$\micron$, and MIPS 24~$\micron$ are presented, respectively, as blue, green, and red.
The contours begin at 10~mJy~beam$^{-1}$ and increase in steps of 5~mJy~beam$^{-1}$.
Cyanopolyyne peak (blue triangle) and ammonia peak (red triangle) are marked. 
The large orange box represents IRAS 04381+2540. 
The small orange  boxes indicate classified YSO candidates and they are numbered in order of increasing right ascension.
The contours at the top of the image are TMC-1C \citep{Schnee07}. The yellow star represents the position of HD 29647, but HD 29647 is not shown in this image because the MIPS 24~$\mu$m band does not cover the source.} \label{fig1}
\end{figure}

Figure~\ref{fig1} shows 1.2~mm dust continuum emission (contours) on top of
the {\it Spitzer} three-color image (4.5, 8.0, and 24~$\micron$ for blue,
green, and red, respectively). 
The 1.2~mm dust continuum emission shows that the contours around the cyanopolyyne peak are complex compared to those around the ammonia peak.
The IRAS source (IRAS 04381+2540) marked
with a large box, is located in the northwest part of TMC-1. The small
boxes represent YSO candidates that are classified in the c2d source
catalogs \citep{Evans07}. The source identification numbers are
assigned with increasing right ascension.  
With the {\it Spitzer} observations,
we confirm that there are no YSO candidates along the densest part of the
TMC-1 filament although 6 YSO candidates (including the IRAS source) are
identified outside or at boundary of the TMC-1 filament.  
There is a gradient in the 4.5 and 8~$\micron$ emission from northwest to southeast direction 
probably due to IRAS 04380+2553, which is a late
B-type star, HD 29647, behind TMC-1 \citep{Whittet04, Mooley13}.
The 4.5 and 8~$\micron$ radiation from HD 29747 is scattered by dust grains in low densities. Therefore, the gradient of the scattered light emission 
might imply that the TMC-1 filament is inclined toward HD 29647. As a result, the surface of its northwest part is more illuminated by the B star.

\begin{figure*}[t!]
\centerline{
\includegraphics[width=3.3in,height=3.0in]{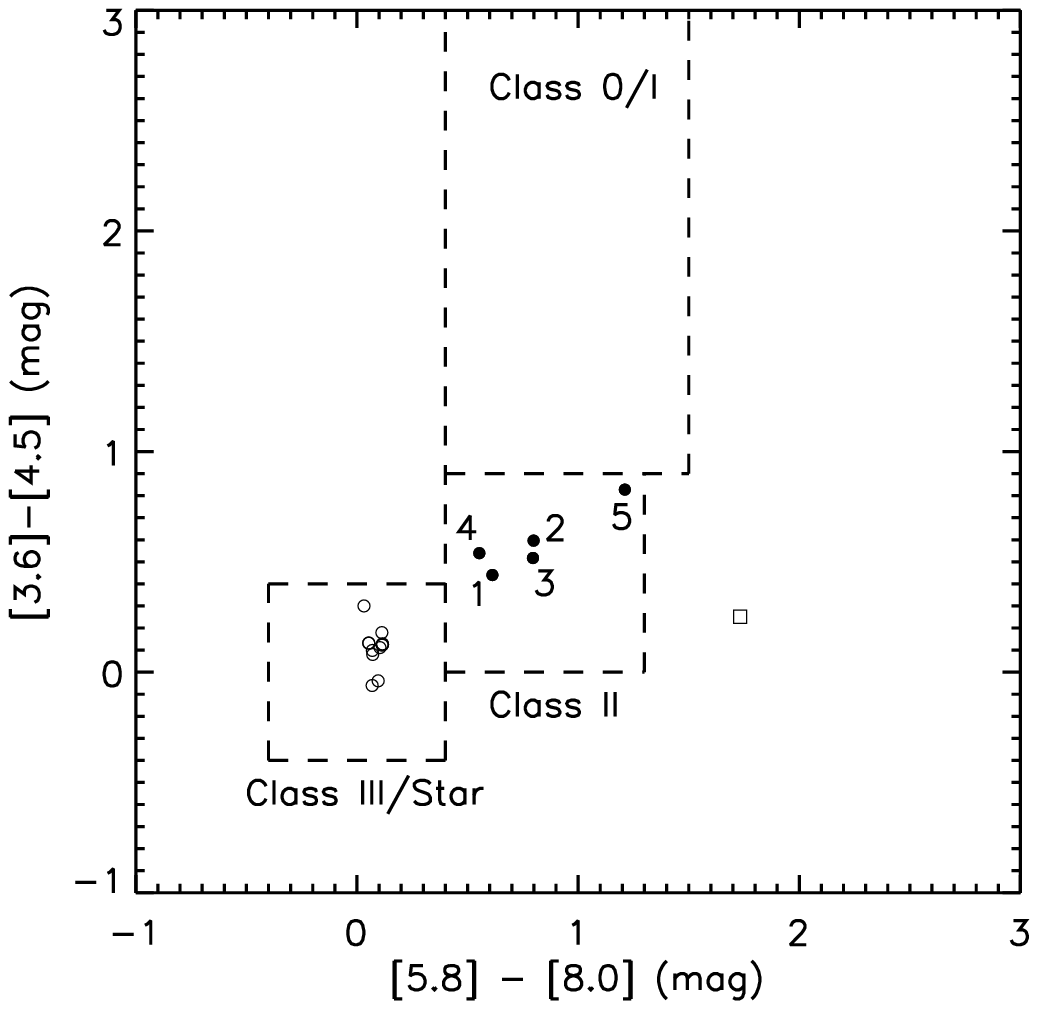}\hspace*{0.009\textwidth}
\includegraphics[width=3.3in,height=3.0in]{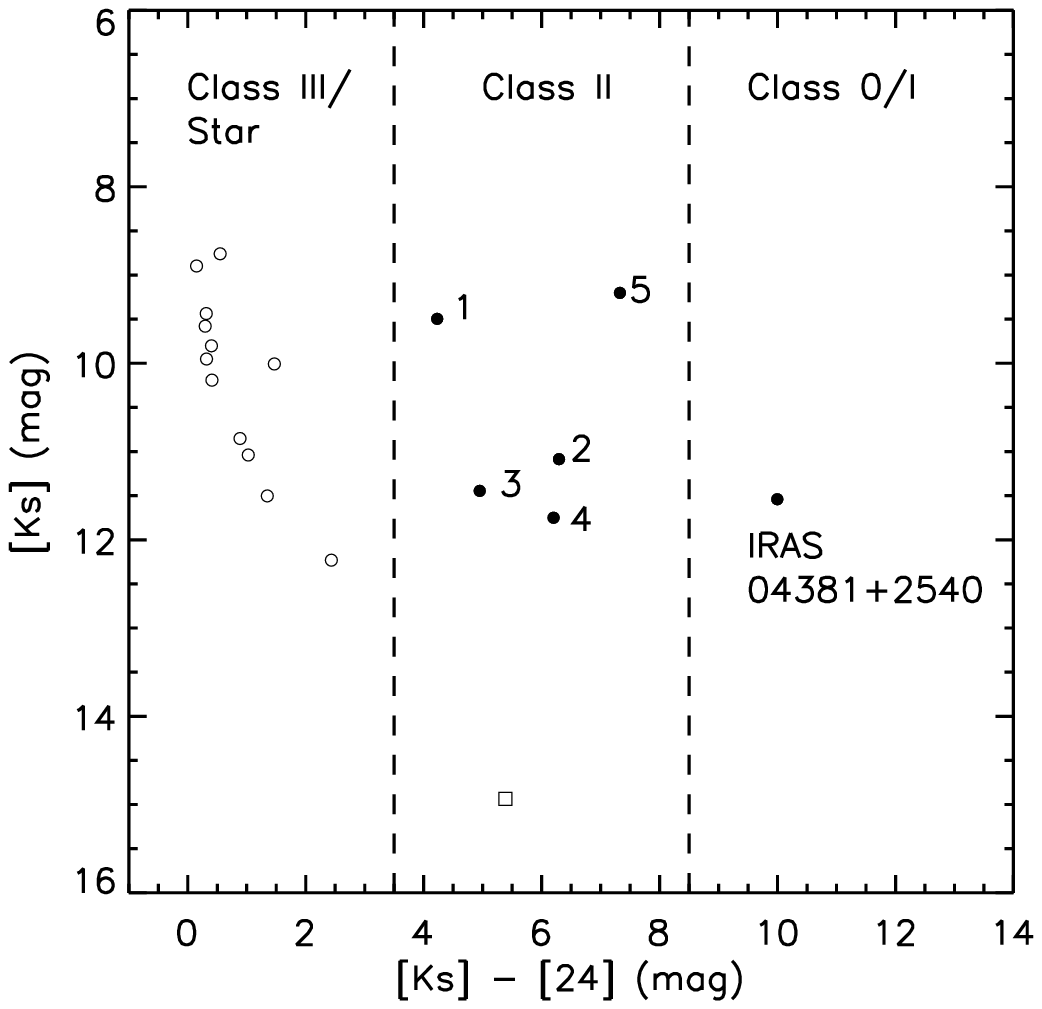}}
\caption{$Left:$ Color-color diagram for sources in the IRAC bands ([3.6]-[4.5]
vs. [5.8]-[8.0]). The filled circles with numbers denote five YSO candidates in TMC-1 (see Fig.~\ref{fig1}). The open circles and open square indicate stars and a galaxy, respectively, based on the c2d classification \citep{Evans07}.
The dashed boxes indicate the approximate domains of Class 0/I, Class II,
and Class III sources as used in \citet{Lee06}. IRAS 04381+2540  is not included because two IRAC bands (3.6 and 5.8~$\micron$) do not cover the source.
$Right:$ Color-magnitude diagram for the 2MASS Ks-band and the MIPS 24~$\micron$ band sources.
This diagram includes IRAS source as well as 5 YSO candidates. The symbols are the same as in the left diagram. The three domains divided by dashed line show different evolutionary stages. \label{fig2}}
\end{figure*}

To check the evolutionary stages of the 6 YSO candidates we use the same
color-color and color-magnitude diagrams (Fig.~\ref{fig2}) as used in
\citet{Lee06}.  
The diagrams use the photometric data from the 2MASS \citep[Two Micron All Sky Survey;][]{Skrutskie06}, IRAC, and MIPS bands.
IRAS 04381+2540 is not included in the color-color diagram
since it is not covered by the observations at 3.6 and 5.8~$\micron$.  
The coordinates and fluxes from the 2MASS and  {\it Spitzer} observations of these YSO candidates are listed in
Table~\ref{tbl-2}.  As seen in Fig.~\ref{fig2}, the IRAS source is a Class
0/I candidate, and 5 YSO candidates are classified as Class II.
We compare our classification to previous results in the literature. 6 YSOs including the IRAS source are in good agreement with previously identified objects in the catalogs of \citet{Gutermuth08, Gutermuth09} and \citet{Rebull10}. Table~\ref{tbl-14} lists previous studies, our classification, and identifications for 6 YSOs.

\tabletypesize{\tiny}
\begin{deluxetable}{llcccccccccccc}
\tablecolumns{14}
\tablewidth{0pc} \setlength{\tabcolsep}{0.02in}
\tablecaption{Fluxes of Young Stellar Object Candidates in TMC-1.\label{tbl-2}}
\tablehead{
\colhead{} & \colhead{} & \colhead{} & \colhead{} & \multicolumn{3}{c}{Fluxes from 2MASS (mJy)} & \colhead{} &\multicolumn{6}{c}{Fluxes from {\it Spitzer} (mJy)}  \\
\cline{5-7} \cline{9-14} 
\colhead{Source\tablenotemark{a}} & \colhead{c2d ID} & \colhead{R.A. (J2000)} & \colhead{Dec. (J2000)} & \colhead{J} & \colhead{H} & \colhead{Ks} & \colhead{} & \colhead{IRAC1} & \colhead{IRAC2} & \colhead{IRAC3} & \colhead{IRAC4} &\colhead{MIPS1} & \colhead{MIPS2}  \\
\colhead{} & \colhead{} & \colhead{($^{\rm h}$ $^{\rm m}$ $^{\rm s}$)} & \colhead{($\arcdeg$ $\arcmin$ $\arcsec$)} & \colhead{(1.235~$\micron$)} & \colhead{(1.662~$\micron$)} & \colhead{(2.159~$\micron$)}  &\colhead{} & \colhead{(3.6~$\micron$)} & \colhead{(4.5~$\micron$)} & \colhead{(5.8~$\micron$)} & \colhead{(8.0~$\micron$)} &\colhead{(24~$\micron$)} & \colhead{(70 $\micron$)}}
\startdata
YSO 1 & SSTc2d J044049.5+255119 & 04 40 49.5 & 25 51 19.1 & 7.99E+01 & 1.10E+02 & 1.06E+02 & & 6.69E+01 & 6.42E+01 & 5.54E+01 & 5.43E+01 & 5.58E+01 & \\
YSO 2 & SSTc2d J044108.2+255607 & 04 41 08.2 & 25 56 07.4 & 5.11E+00 & 1.42E+01 & 2.45E+01 & & 3.71E+01 & 4.11E+01 & 4.46E+01 & 5.19E+01 & 8.64E+01  & \\
YSO 3 & SSTc2d J044110.8+255512 & 04 41 10.8 & 25 55 11.6 & 8.42E+00 & 1.45E+01 & 1.76E+01 & & 1.32E+01 & 1.36E+01 & 1.37E+01 & 1.59E+01 & 1.80E+01  &\\
IRAS source\tablenotemark{b} & SSTc2d J044112.6+254635 &  04 41 12.6 & 25 46 35.4 & 2.20E-01 & 2.27E+00 & 1.61E+01 & & & 1.39E+02 & & 3.46E+02 & 1.72E+03 & 6.74E+03  \\
YSO 4 & SSTc2d J044124.6+254353 & 04 41 24.6 & 25 43 53.0 & 3.30E-01 & 3.95E+00 & 1.33E+01 & & 1.95E+01 & 2.05E+01 & 2.09E+01 & 1.94E+01 & 4.31E+01  & \\
YSO 5 & SSTc2d J044138.8+255627 & 04 41 38.8 & 25 56 26.8 & 2.90E+01 & 9.14E+01 & 1.39E+02 & & 1.16E+02 & 1.59E+02 & 2.48E+02 & 4.22E+02 & 1.27E+03 & 2.11E+03  \\
\enddata
\tablenotetext{a}{The numbers refer to the marked YSOs in Fig.~\ref{fig1}.}
\tablenotetext{b}{IRAS 04381+2540.}
\end{deluxetable}

We also calculate the bolometric luminosity ($L_{\rm bol}$) and bolometric temperature ($T_{\rm bol}$) for these 6 YSO candidates using the observed fluxes (see Table~\ref{tbl-2}).
The bolometric luminosity ($L_{\rm bol}$) is calculated by integrating the spectral energy distributions (SEDs), including far-infrared and sub-millimeter wavelengths by extrapolation.

\begin{equation}\label{eq1}
L_{\rm bol} = 4\pi d^2\int_{0}^{\infty}{S_{\nu}d\nu},
\end{equation}
where $d$ is the distance to the source and $S_{\nu}$ is the flux as a function of frequency of the source.
The bolometric temperature ($T_{\rm bol}$) is described as the temperature of a blackbody with the same flux-weighted mean frequency as the observed SED and it is calculated using the following equation \citep{Myers93}:
\begin{equation}\label{eq2}
T_{\rm bol} = 1.25\times10^{-11}\frac{\int_{0}^{\infty} \nu S_{\nu} d\nu}{\int_{0}^{\infty} S_{\nu} d\nu} \,\,\,\,\, {\rm K}.
\end{equation}

Table~\ref{tbl-3} shows that the evolutionary stage of each
source based on the bolometric temperature 
\citep[Class I : 70~K $\leq$ $T_{\rm bol}$ $\leq$ 650~K, Class II : 650~K $<$ $T_{\rm bol}$ $\leq$ 2880~K;][]{Chen95}
is consistent with the result found using the color-color and color-magnitude diagrams. 
Furthermore, \citet{Dunham08} calculated the bolometric luminosity and temperature of $\sim$0.73~L$_\sun$ and $\sim$165~K for the IRAS source, which is also classified as a Class I object.

\tabletypesize{\small}
\begin{deluxetable}{llcccc}[ht!]
\tablecolumns{6}
\tablewidth{0pt}
\tablecaption{Comparison with Previous Studies. \label{tbl-14}}
\tablehead{
\colhead{Source\tablenotemark{a}} & \colhead{c2d ID} & \colhead{Gutermuth Category\tablenotemark{b}} & \colhead{Rebull Category\tablenotemark{c}} & \colhead{This Study} & \colhead{Identification\tablenotemark{d}}}
\startdata
YSO 1 & SSTc2d J044049.5+255119 & Class II &  Class II & II & JH 223\\
YSO 2 & SSTc2d J044108.2+255607 & Class II &  Flat & II & ITG 33A\\
YSO 3 & SSTc2d J044110.8+255512 & Class II &  Class II & II & ITG 34\\
IRAS source & SSTc2d J044112.6+254635 & Embedded & Class I & 0/I & IRAS 04381+2540  \\
YSO 4 & SSTc2d J044124.6+254353 & Class II & Flat & II  & GKH 32 \\
YSO 5 & SSTc2d J044138.8+255627 & Class II & Flat & II  & Haro 6-33 \\
\enddata
\tablenotetext{a}{The numbers refer to the marked YSOs in Fig.~\ref{fig1}.}
\tablenotetext{b}{\citet{Gutermuth08, Gutermuth09}}
\tablenotetext{c}{\citet{Rebull10}}
\tablenotetext{d}{Reference for the first identification of each source: JH 223 - \citet{Jones79}; ITG 33A, ITG 34 - \citet{Itoh99}; IRAS 04381+2540 - \citet{Beichman86}; GKH 32 - \citet{Gomez94}; Haro 6-33 - \citet{Haro53}.}
\end{deluxetable}

\clearpage
\tabletypesize{\small}
\begin{deluxetable*}{llcccc}[ht!]
\tablecolumns{6}
\tablewidth{0pt}
\tablecaption{Results of Bolometric Luminosity and Temperature of 6 YSO Candidates in TMC-1.\label{tbl-3}}
\tablehead{
\colhead{Source\tablenotemark{a}} & \colhead{c2d ID} & \colhead{$L_{\rm bol}$ ($L_\sun$)} &
\colhead{} & \colhead{$T_{\rm bol}$ (K)} & \colhead{CLASS}}
\startdata
YSO 1 & SSTc2d J044049.5+255119 & 0.12 &  & 1722 & II \\
YSO 2 & SSTc2d J044108.2+255607 & 0.04 &  & 991 & II \\
YSO 3 & SSTc2d J044110.8+255512 & 0.02 &  & 1460 & II \\
IRAS source\tablenotemark{b} & SSTc2d J044112.6+254635 & 0.54 &  & 214 & I \\
YSO 4 & SSTc2d J044124.6+254353 & 0.02 &  & 922  & II \\
YSO 5 & SSTc2d J044138.8+255627 & 0.33 &  & 758  & II \\
\enddata
\tablenotetext{a}{The numbers refer to the marked YSOs in Fig.~\ref{fig1}.}
\tablenotetext{b}{IRAS 04381+2540.}
\end{deluxetable*}

\subsection{Molecular Line Observations}

\begin{figure}[t!]
\epsscale{0.6}
\plotone{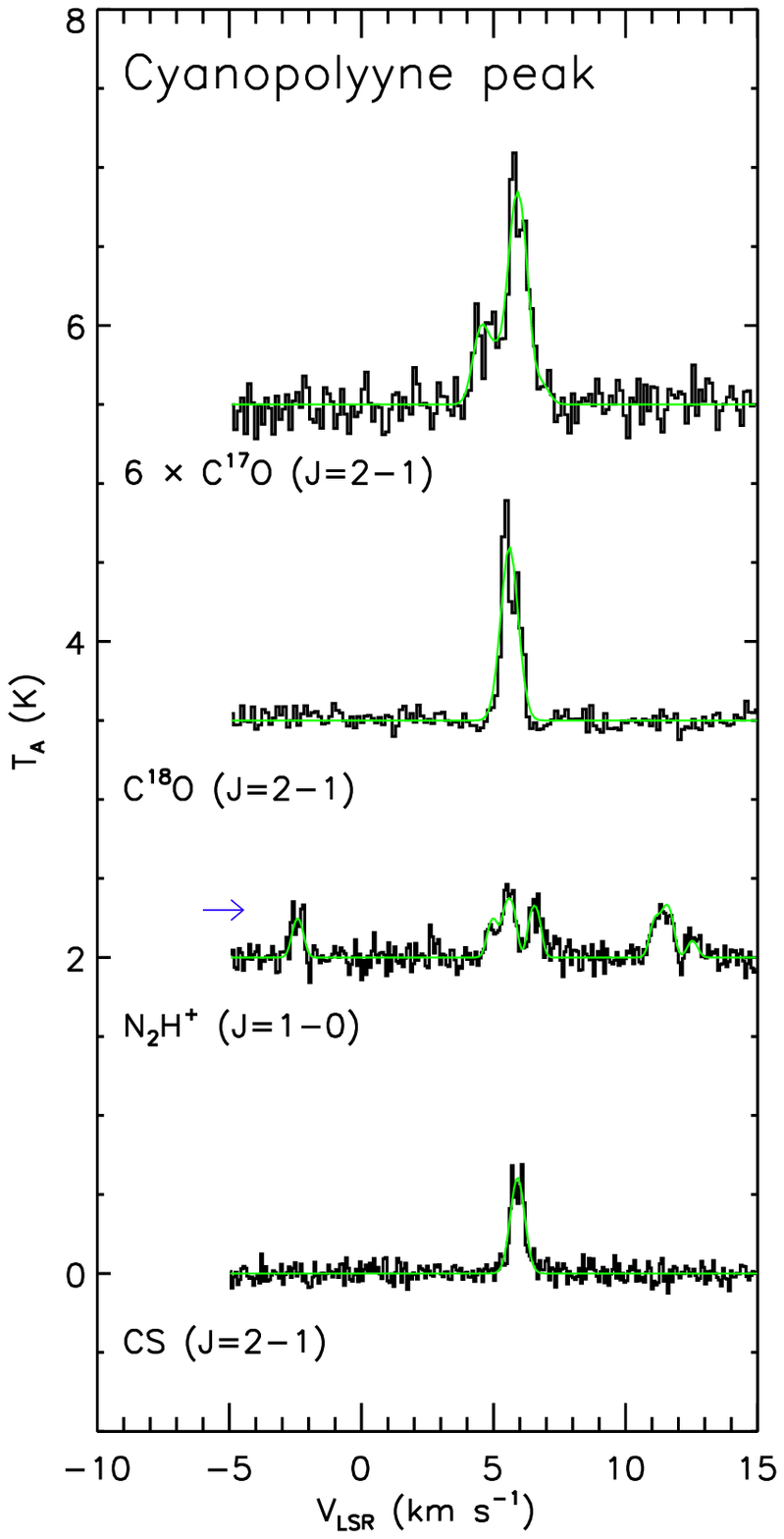}
\caption{Spectra of molecular transitions observed with the FCRAO 14~m and the SRAO 6~m
telescopes at the cyanopolyyne peak (black). The gaussian-fitting results and hyperfine-fitting results for CS 2--1/C$^{18}$O 2--1 and N$_2$H$^+$ 1--0/C$^{17}$O 2--1 are shown as green lines, respectively. The isolated component of N$_2$H$^+$ 1--0 is pointed out by the blue arrow.} \label{fig4}
\end{figure}

\begin{figure}[t!]
\epsscale{0.6}
\plotone{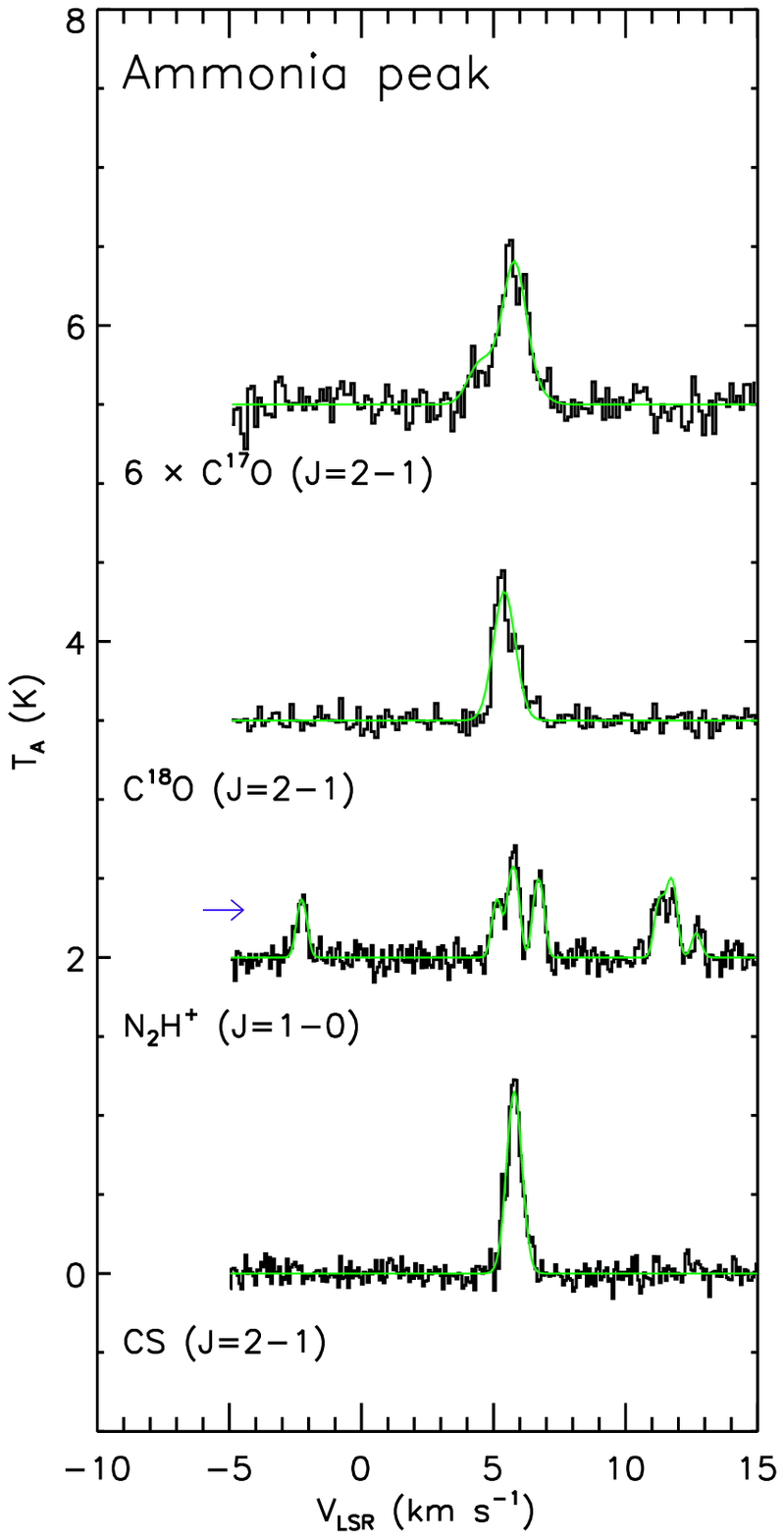}
\caption{Spectra of molecular transitions observed with the FCRAO 14~m and the SRAO 6~m
telescopes at the ammonia peak. The gaussian-fitting results and hyperfine-fitting results for CS 2--1/C$^{18}$O 2--1 and N$_2$H$^+$ 1--0/C$^{17}$O 2--1 are shown as green lines, respectively. The isolated component of N$_2$H$^+$ 1--0 is pointed out by the blue arrow.} \label{fig5}
\end{figure}

Figures~\ref{fig4} and~\ref{fig5} present four spectra observed at the cyanopolyyne
and ammonia peaks. The Gaussian fitting results of CS 2$-$1 and C$^{18}$O 2--1 are obtained using the GILDAS/CLASS package\footnote{\url{http://www.iram.fr/IRAMFR/GILDAS/}} and they are summarized in Table~\ref{tbl-5}.
N$_2$H$^+$ 1--0 and C$^{17}$O 2--1 have hyperfine structures and they are fitted using the method in the program CLASS (METHOD HFS), 
which fits the relative intensities of the hyperfine structure lines of a molecular rotational transition to derive the optical depth of the line.
The hyperfine fitting results are listed in Table~\ref{tbl-6}. 
CS 2--1, C$^{18}$O 2--1, and the isolated component of N$_2$H$^+$ 1--0, especially at the cyanopolyyne peak, show double peaks.
This double peak feature could be caused either by optical depth or by
two velocity components. Here, we assume that the line shape is caused
by optical depth, rather than two velocity components, and fit the data
with a single Gaussian.

\begin{figure}
\epsscale{0.8}
\plotone{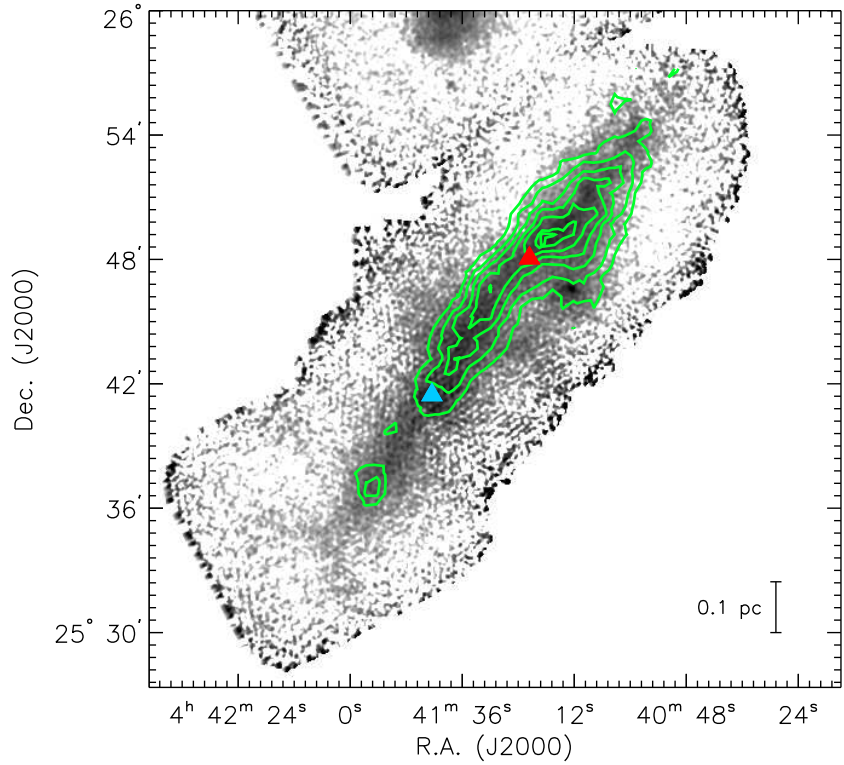}
\caption{Integrated intensity maps of CS 2--1 (green contours)
on top of the 1200~$\micron$ dust continuum emission map (gray scale).
Blue and red triangles denote the cyanopolyyne peak and the ammonia peak, respectively.
The contour values are 0.38, 0.53, 0.68, 0.83, 0.98, 1.13, 1.17~K~km~s$^{-1}$. \label{fig6}}
\end{figure}

\begin{figure}
\epsscale{0.8}
\plotone{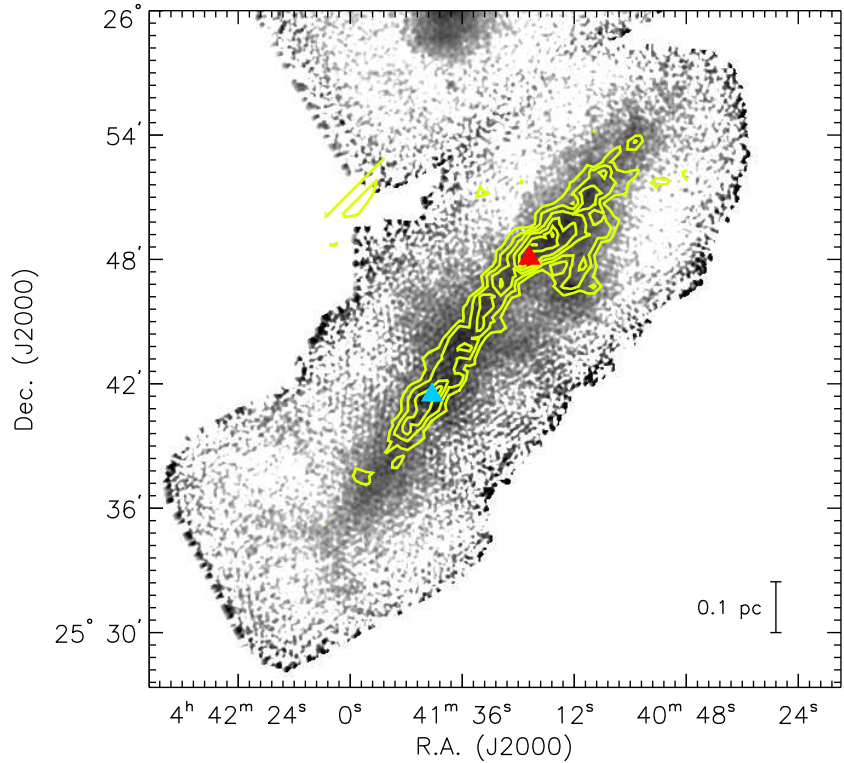}
\caption{Integrated intensity maps of N$_2$H$^+$ 1--0 (yellow contours)
on top of the 1200~$\micron$ dust continuum emission map (gray scale).
Blue and red triangles denote the cyanopolyyne peak and the ammonia peak, respectively.
The contour values are 0.08, 0.12, 0.16, 0.20, 0.24, 0.27~K~km~s$^{-1}$. \label{fig7}}
\end{figure}

Figures~\ref{fig6} and~\ref{fig7} present the integrated intensity maps of CS 2--1 (green line) and N$_2$H$^+$ 1--0 (yellow line) obtained at the FCRAO. In the figure, line emissions (contours) are overlaid on the 1.2~mm dust continuum emission map (grey image), and the locations of the cyanopolyyne peak and the ammonia peak are denoted by red and blue triangles, respectively.
This 1.2~mm dust continuum emission map, the best tracer of column density, is the first 1.2~mm dust continuum emission map of its type for the whole TMC-1 region \citep{Kauffmann08}. The CS 2--1 intensity does not peak at the cyanopolyyne peak, while it peaks close to the ammonia peak. 
However, the N$_2$H$^+$ 1--0 intensity peaks around both cyanopolyyne and ammonia peaks
although the intensity close to the cyanopolyyne peak is weaker than that around the ammonia peak.
The differences of the molecular line intensities at the two peaks 
can be caused by differences in
(1) density, (2) temperature, (3) abundance, or any 
combinations of these properties.

\begin{figure}
\epsscale{0.8}
\plotone{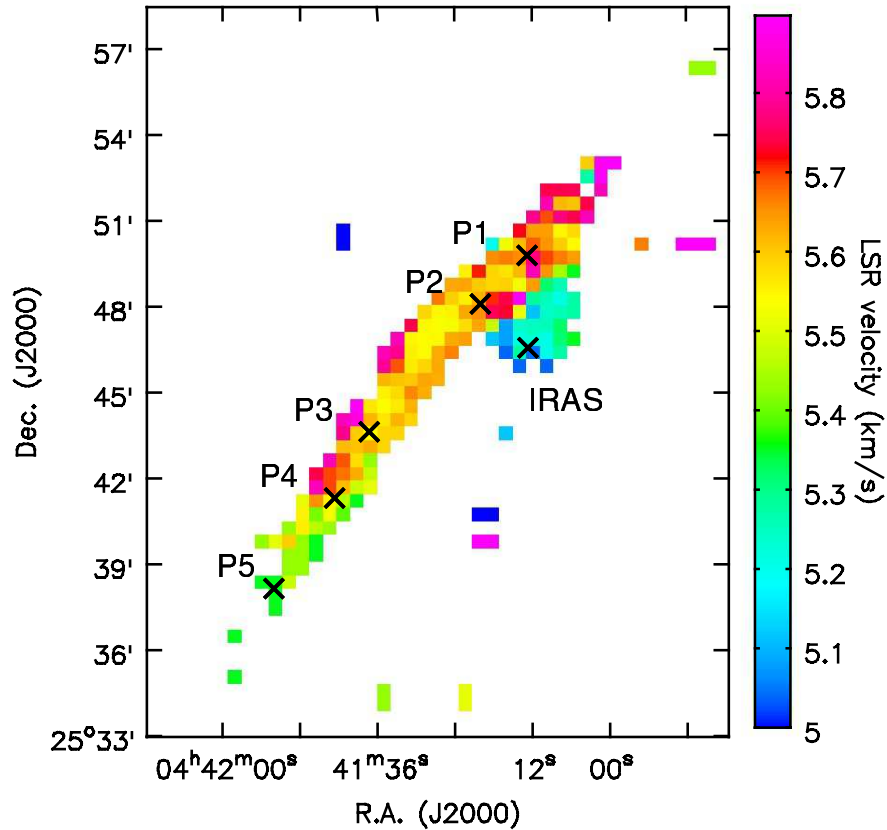}
\caption{The first moment (i.e., intensity weighted central velocity) map of the N$_2$H$^+$ 1--0 line. Five positions (P1, P2, P3, P4, and P5) are placed along the TMC-1 filament in order of increasing right ascension. The IRAS source (IRAS 04381+2540) is shown in the northwest part of the TMC-1 filament. The coordinate and $V_{\rm LSR}$ of each position is listed in Table~\ref{tbl-7}.} \label{fig8}
\end{figure}

We use the N$_2$H$^+$ 1--0 line to study the velocity field because that
line has the lowest optical depth. Figure~\ref{fig8} presents the first moment map. 
We found three velocity components; the northern part (red to yellow), the southern part (green), and the northwestern part (sky blue). 
In the northern part (P1) of the TMC-1 filament, the velocity is $\sim$5.8~km~s$^{-1}$ and it drops to about 5.4~km~s$^{-1}$ in the  the southern part (P5). 
On the other hand, the source velocity in the northwestern part (IRAS 04381+2540, $V_{\rm LSR}\sim$5.2~km~s$^{-1}$)  is lower than the typical velocity along the TMC-1 filament ($V_{\rm LSR}\sim$5.6~km~s$^{-1}$), suggesting that this region may be separated from the TMC-1 filament along the line of sight \citep{Toelle81, Snell82, Kolotovkina86, Olano88, Feher16}.

Table~\ref{tbl-7} summarizes the source velocities of five positions (P1, P2, P3, P4, and P5) along the TMC-1 filament, consistent with the MAMBO dust continuum peaks \citep{Kauffmann08}, and the IRAS source (IRAS 04381+2540).

\clearpage

\clearpage

\begin{deluxetable*}{lccccccccc}
\tabletypesize{\small}
\tablecolumns{10}
\tablewidth{0pt}
\tablecaption{Results of Gaussian fits to CS ($J$=2--1) and C$^{18}$O ($J$=2--1) Lines.\label{tbl-5}}
\tablehead{
\colhead{} & \multicolumn{4}{c}{Cyanopolyyne Peak} & \colhead{} & \multicolumn{4}{c}{Ammonia Peak} \\
\cline{2-5} \cline{7-10} 
\colhead{Line} & \colhead{$V_{\rm LSR}$} & \colhead{$\Delta V$} & \colhead{$T_{\rm A}$} &
\colhead{$\int T_{\rm A}dV$} & \colhead{} & \colhead{$V_{\rm LSR}$} & \colhead{$\Delta V$} & \colhead{$T_{\rm A}$} &
\colhead{$\int T_{\rm A}dV$} \\
\colhead{} & \colhead{(km s$^{-1}$)} & \colhead{(km s$^{-1}$)} & \colhead{(K)} &
\colhead{(K km s$^{-1}$)} & \colhead{} & \colhead{(km s$^{-1}$)} & \colhead{(km s$^{-1}$)} & \colhead{(K)} &
\colhead{(K km s$^{-1}$)}}
\startdata
CS ($J$=2--1)  & 5.91   & 0.63   & 0.60 & 0.40   & & 5.78   & 0.70   & 1.16 & 0.86 \\
                        & (0.01) & (0.03) &       & (0.01) & & (0.01) & (0.02) &       & (0.02) \\
C$^{18}$O ($J$=2--1)   & 5.61   & 0.76   & 1.08 & 0.87   & & 5.42   & 0.94   & 0.80 & 0.80 \\
                                     & (0.01) & (0.02) &       & (0.01) & & (0.02) & (0.05) &       & (0.03) \\
\enddata
\tablecomments{The values in parentheses refer to errors of each parameter.}
\end{deluxetable*}

\begin{deluxetable*}{lccccccc}[h!]
\tablecolumns{8}
\tablewidth{0pt}
\tablecaption{Results of Hyperfine structure to N$_2$H$^+$ ($J$=1--0) and C$^{17}$O ($J$=2--1) Lines.\label{tbl-6}}
\tablehead{
\colhead{} & \multicolumn{3}{c}{Cyanopolyyne Peak} & \colhead{} & \multicolumn{3}{c}{Ammonia Peak} \\
\cline{2-4} \cline{6-8} 
\colhead{Line} & \colhead{$V_{\rm LSR}$} & \colhead{$\Delta V$} & \colhead{$\tau$\tablenotemark{a}} &
\colhead{} & \colhead{$V_{\rm LSR}$} & \colhead{$\Delta V$} & \colhead{$\tau$\tablenotemark{a}}  \\
\colhead{} & \colhead{(km s$^{-1}$)} & \colhead{(km s$^{-1}$)} & \colhead{} &
\colhead{} & \colhead{(km s$^{-1}$)} & \colhead{(km s$^{-1}$)} & \colhead{}}
\startdata
N$_2$H$^+$ ($J$=1--0) & 5.60     & 0.44    & 7.64    & & 5.75       & 0.41      & 6.77   \\
                                        & (0.01) & (0.03)  & (2.11)  & & (0.01) & (0.02) & (1.63)  \\
C$^{17}$O ($J$=2--1)    & 5.65       & 0.68       & 1.71         & & 5.53       & 1.01       & 0.10   \\
                                          & (0.02) & (0.05) & (0.66)  & & (0.02) & (0.04) & (0.16)  \\
\enddata
\tablecomments{The values in parentheses refer to errors of each parameter.}
\tablenotetext{a}{$\tau$ is the sum of the optical depths of all hyperfine components.}
\end{deluxetable*}

\begin{deluxetable*}{lccc}
\tablecolumns{4}
\tablewidth{0pt}
\tablecaption{LSR Velocity in the TMC-1 filament.\label{tbl-7}}
\tablehead{
\colhead{Position} & \colhead{R.A. (J2000)} & \colhead{Dec. (J2000)} & \colhead{LSR Velocity} \\
\colhead{} &  \colhead{($^{\rm h}$ $^{\rm m}$ $^{\rm s}$)}  & \colhead{($\arcdeg$ $\arcmin$ $\arcsec$)} & \colhead{(km s$^{-1}$)}
}
\startdata
\multicolumn{4}{c}{The direction along the TMC-1 filament}\\
\cline{1-4}
P1                     & 04 41 13.8 & 25 49 49.5 & 5.82 (0.01) \\
P2\tablenotemark{a}                     & 04 41 20.9 & 25 48 07.2 & 5.75 (0.01) \\
P3                     & 04 41 37.2 & 25 43 38.5 & 5.66 (0.01)\\
P4\tablenotemark{b}                     & 04 41 42.5 & 25 41 27.0 & 5.60 (0.01) \\
P5                     & 04 41 51.9 & 25 38 09.5 & 5.42 (0.01) \\
\cline{1-4}
IRAS\tablenotemark{c}  & 04 41 12.8 & 25 46 33.5 & 5.20 (0.01) \\
\enddata
\tablecomments{The values in parentheses refer to errors of LSR velocities.}
\tablenotetext{a}{Ammonia peak.}
\tablenotetext{b}{Cynaopolyyne peak.}
\tablenotetext{c}{IRAS 04381+2540.}
\end{deluxetable*}

\clearpage
\section{Column Density Calculations}
\label{sect-4}
\subsection{H$_2$ Column Density from Dust Continuum Emission}
We compare H$_2$ column densities calculated from molecular line emission and dust continuum emission to calculate the degree of depletion of molecules.

If the dust continuum emission is optically thin, the observed flux ${F}^{\rm beam}_{\nu}$
can be related to the column density of gas.
Therefore, we calculate the H$_2$ column density with the MAMBO flux as given by this equation \citep{Hildebrand83}.
\begin{equation}\label{eq3}
N({\rm H_2}) = \frac{{F}^{\rm beam}_{\nu}}{\Omega_{\rm A}\mu_{\rm{H_{2}}}m_{\rm H}\kappa_{\nu}B_{\nu}(T)}
\end{equation}
where ${F}^{\rm beam}_{\nu}$ is the observed flux per beam, $\mu_{\rm{H_{2}}}$ (= 2.3)
is the molecular weight per hydrogen molecule,
$m_{\rm H}$ is the atomic mass unit, $\kappa_{\nu}$ (= 0.0102~cm$^2$g$^{-1}$)
is the mass opacity of dust per gram of gas at 1.2~mm \citep{Kauffmann08},
$B_{\nu}(T)$ is the Planck function, and $\Omega_{\rm A}$ is the beam solid angle;
$\Omega_{\rm A}=(\pi\theta^{2}_{\rm HPBW})/(4\rm{ln}2)$ for a Gaussian function
\citep{Lee03, Kauffmann08}.
The MAMBO map is convolved with a Gaussian profile with the FWHM of the beam of molecular observations.
As a result, the H$_2$ column density calculated from dust continuum emission at the cyanopolyyne and the ammonia peak is (1.99$\pm$0.20)$\times$10$^{22}$~cm$^{-2}$ and (1.87$\pm$0.19)$\times$10$^{22}$~cm$^{-2}$, respectively. The H$_2$ column densities calculated from the dust continuum emission, the best tracer of the column density along the line of sight, is essentially identical at the two peaks.

Two other recent studies also concluded that the column densities in the ammonia and cyanopolyyne peaks are comparable. \citet{Suutarinen11} estimated the total H$_2$ column density in the TMC-1 region using the SCUBA 850 and 450~$\micron$ data from \citet{Nutter08}. The derived $N$(H$_2$) are $\sim1.3\times10^{22}$~cm$^{-2}$ and $\sim1.6\times10^{22}$~cm$^{-2}$ at the ammonia peak and the cyanopolyyne peak, respectively.  
In addition, \citet{Feher16} used the {\it Herschel}/SPIRE images to derive the column densities of $3.3\times10^{22}$~cm$^{-2}$ at the ammonia peak and $2.7\times10^{22}$~cm$^{-2}$ at the cyanopolyyne peak. These absolute values are different from ours probably because of different beam sizes. However, the very important point of these calculations including ours is that the column densities at the ammonia and cyanopolyyne peaks are 
similar, contrary to suggestions by some previous studies (see Section~\ref{sect-1}).

\subsection{Column Density and Abundance for Molecular Line Emission}
Chemical models predict that CO and N$_2$H$^+$ will be  
anti-correlated because CO destroys N$_2$H$^+$ in the gas phase. 
Previous observations of pre- and protostellar regions 
\citep{Bergin01, Tafalla02, Lee03, DiFrancesco04, Jorgensen04, Pagani05} 
confirmed that N$_2$H$^+$ becomes abundant as CO is frozen onto grain surfaces. Therefore, these two molecules together can trace different chemical evolutionary stages of molecular cores.
We use the N$_2$H$^+$ and C$^{17}$O lines to calculate column densities more accurately
because these lines have hyperfine structures.
In this calculation, we assume that (1) the dust continuum emission traces all
material along a line of sight; (2) dust and gas are well coupled thermally \citep[$T_{\rm d}$ = $T_{\rm k}$ = 10~K,][]{Pratap97}; (3) molecular abundances are constant along the line of sight;
and (4) all levels are in LTE, following \citet{Lee03}.

For line emission, if a line is optically thin ($\tau$ $<$ 0)
for the transition $J=J\rightarrow J-1$
in a linear molecule, the relation between
the column density $N(x)$ of molecule $x$ and the integrated intensity of the line is

\begin{equation}\label{eq4}
N(x) = \frac{3kQe^{E_{J}/kT_{\rm ex}}}{8\pi^{3}\nu\mu^{2}J}\int{T_{R}dv}
\end{equation}
where $T_{\rm ex}$ is the excitation temperature
above the ground state, and $\mu$ is the dipole moment. 
The frequency ($\nu$), energy above ground ($E_{J}$), and 
the partition function ($Q$) can be computed from $B$, 
the rotational constant.

\tabletypesize{\small}
\begin{deluxetable*}{lcc}[t!]
\tablecolumns{3}
\tablewidth{0pt}
\tablecaption{Parameters for Molecules.\label{tbl-8}}
\tablehead{
\colhead{Line} & \colhead{Rotational Constant B}   & \colhead{Dipole Moment $\mu$}  \\
\colhead{} & \colhead{(MHz)} & \colhead{(D\tablenotemark{a})}}
\startdata
N$_2$H$^+$ 1--0 & 46586.87 & 3.40   \\
C$^{17}$O 2--1 &  56179.99 & 0.11   \\
\enddata
\tablecomments{Molecular spectroscopy data are from the CDMS catalog \citep[][\url{http://www.astro.uni-koeln.de/cdms}]{Muller05}.}
\tablenotetext{a}{The Debye (D) is a unit of the electric dipole moment, 1~D = 10$^{-18}$~g$^{1/2}$~cm$^{5/2}$~s.}
\end{deluxetable*}

However, if the optical depth is not negligible, the optical depth effect is corrected for
with the equation :
\begin{equation}\label{eq5}
N_{\rm thick} = N_{\rm thin}\frac{\tau_{\nu}}{1-e^{-\tau_{\nu}}}
\end{equation}
Additionally, we calculate the abundance of molecule, $X(x)$, using the H$_2$ column density calculated from the dust continuum emission.
\begin{equation}\label{eq6}
X(x)=\frac{N(x)}{N(\rm H_2)_{\rm Dust}}
\end{equation}

Table~\ref{tbl-8} summarizes the parameters (the rotational constant $B$ and the dipole moment $\mu$) used for the calculation of the column densities.
The optical depths of the N$_2$H$^+$ 1--0 and C$^{17}$O 2--1 lines are
obtained by the hyperfine fitting method in the CLASS package (see Table~\ref{tbl-6}). 
We find that C$^{17}$O 2--1 is optically thick at the cyanopolyyne peak, but optically
thin at the ammonia peak.
On the other hand, the total optical depth of N$_2$H$^+$ is too large at both peaks to adopt the Eq.~\ref{eq5}. Therefore, we use the isolated component (see Fig.~\ref{fig4} and~\ref{fig5}) to derived the column density of N$_2$H$^+$ by assuming that the optical depth of each component is proportional to its LTE intensity ratio. The optical depth of the isolated component is 10\% of the total optical depth.

The column densities and abundances for molecular lines
are listed in Table~\ref{tbl-10}.
The column density from C$^{17}$O 2--1 is a factor of 2 larger at the cyanopolyyne peak
than at the ammonia peak. However, there is no significant difference
in the column densities of N$_2$H$^+$ 1--0 between the cyanopolyyne and the ammonia peaks
when the error is considered.

Table~\ref{tbl-11} represents the depletion factors of C$^{17}$O and N$_2$H$^+$
when we compare the calculated abundances with the canonical abundances,
$X(\rm {C^{17}O})=1.5\times10^{-7}$ and $X(\rm{N_2H^+})=5.0\times10^{-10}$.
This canonical abundance of C$^{17}$O is obtained by 
assuming $^{16}$O/$^{18}$O of 540, $^{18}$O/$^{17}$O of 3.65 \citep{Wilson94, Penzias81}, and CO/H$_2$ of 2.7$\times$10$^{-4}$ \citep{Lacy94}. The canonical abundance of N$_2$H$^+$ is adopted from \citet{Johnstone10}. 
We find that the depletion of CO is more significant in the ammonia peak (CO depletion factor = 11.9) compared to the cyanopolyyne peak (CO depletion factor = 4.5). This suggests that the ammonia peak is more chemically evolved than the cyanopolyyne peak.
On the other hand, the depletion of N$_2$H$^+$ is almost identical in the two peaks if errors are considered.

\begin{deluxetable*}{lccccc}[t!]
\tabletypesize{\small}
\tablecolumns{6}
\tablewidth{0pt}
\tablecaption{Column Densities and Abundances for Molecular Lines.\label{tbl-10}}
\tablehead{
\colhead{} & \multicolumn{2}{c}{Cyanopolyyne Peak} & \colhead{} & \multicolumn{2}{c}{Ammonia Peak} \\
\cline{2-3}  \cline{5-6} 
\colhead{Line} & \colhead{Column Density} & \colhead{Abundance} & \colhead{} & \colhead{Column Density} & \colhead{Abundance} \\
\colhead{} & \colhead{($\times$10$^{13}$~cm$^{-2}$)} & \colhead{($\times$10$^{-9}$)} &  \colhead{} & \colhead{($\times$10$^{13}$~cm$^{-2}$)} & \colhead({$\times$10$^{-9}$})}
\startdata
N$_2$H$^+$ 1--0 & 0.5 (0.2) & 0.3 (0.1) &   & 0.6 (0.2) & 0.3 (0.1) \\
C$^{17}$O 2--1   & 66.5 (26.7) & 33.4 (13.9)  &  & 25.1 (3.1) & 13.4 (2.0) \\
\enddata
\tablecomments{We correct for the optical depth effect to calculate the column density from N$_2$H$^+$ 1--0 and C$^{17}$O 2--1. The values in parentheses refer to errors of each parameter.}
\end{deluxetable*}

\begin{deluxetable*}{lcc}[ht!]
\tabletypesize{\small}
\tablecolumns{3}
\tablewidth{0pt}
\tablecaption{Depletion Factors of CO and N$_2$H$^+$.\label{tbl-11}}
\tablehead{
\colhead{} & \multicolumn{2}{c}{Depletion Factors ($X_{\rm canonical}$/$X_{\rm measured}$)} \\
\cline{2-3} 
\colhead{} & \colhead{Cyanopolyyne Peak} &  \colhead{Ammonia Peak}}
\startdata
 \decimals
C$^{17}$O & 4.5 (1.9) &  11.9 (1.9) \\
N$_2$H$^+$ & 1.8 (0.7) &  1.6 (0.6) \\
\enddata
\tablecomments{The values in parentheses refer to errors of depletion factors.}
\end{deluxetable*}

\section{Chemical Modeling}
\label{sect-5}

We calculate the evolution of chemical abundances of CO and N$_2$H$^+$ molecules
to explain the differential chemical distribution along the TMC-1 filament.
According to our analysis in the previous section,
the cyanopolyyne and ammonia peaks have similar column densities calculated from the dust continuum emission
while CO is depleted more significantly at the ammonia peak than at the cyanopolyyne peak.
This trend implies that longer timescales at lower densities are necessary at the ammonia peak during the density evolution 
since in that condition, N$_2$ forms more in the gas and CO is frozen-out more onto grain surfaces,
resulting in a high abundance of N$_2$H$^+$ with a significant depletion of CO.
Therefore, we calculate the chemical evolution by varying the 
timescales over which density evolves to account for the observational results.

\begin{figure*}[ht!]
\centerline{\includegraphics[width=4.8in,height=3.8in]{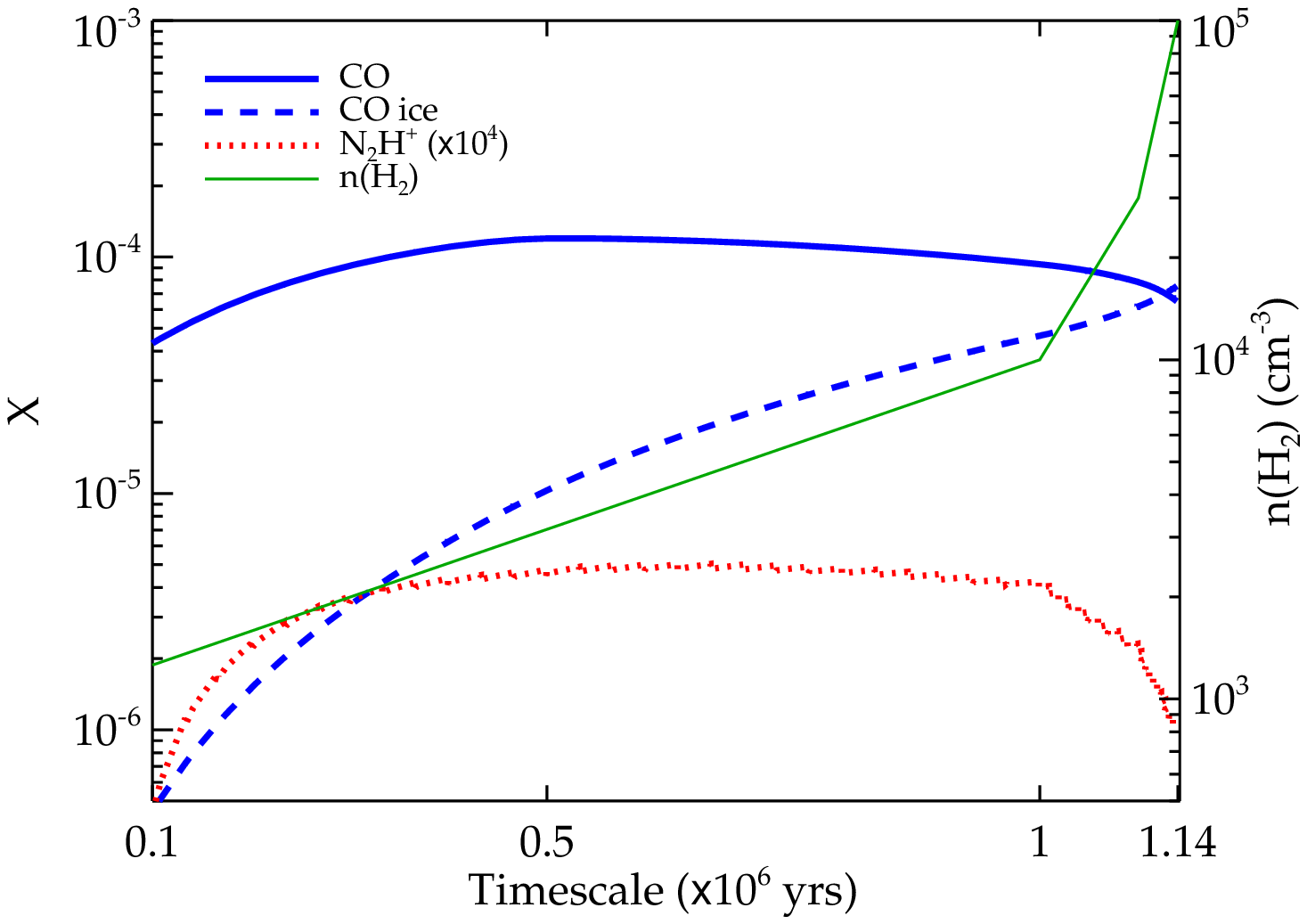}}
\centerline{\includegraphics[width=4.8in,height=3.8in]{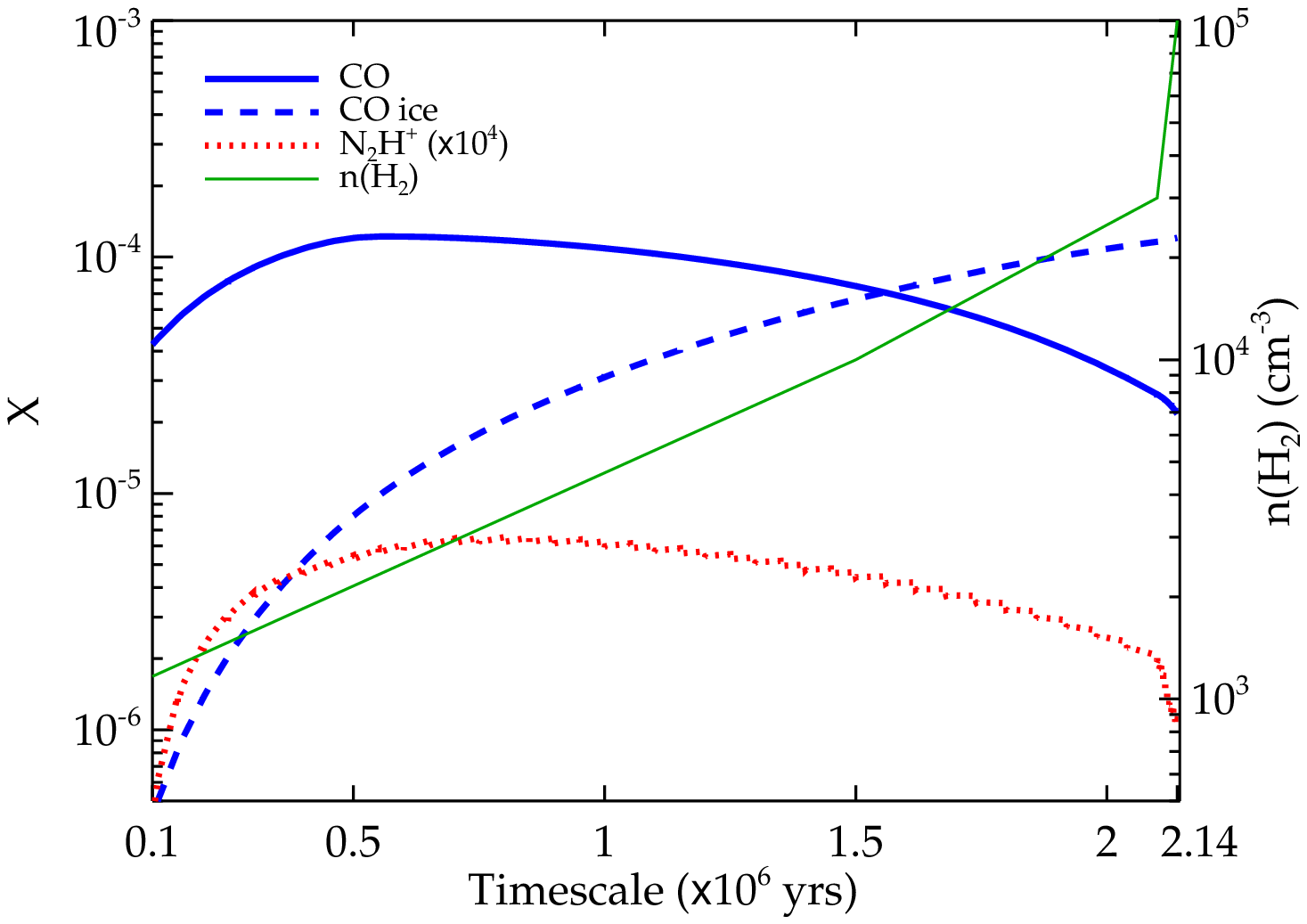}}
\caption{Abundance evolution of CO and N$_2$H$^+$ in the cyanopolyyne peak model ($upper$) and the ammonia peak model ($lower$) from the chemical calculations. The blue solid and dotted lines represent CO abundances in gas and ice, respectively. The red dotted line indicates the N$_2$H$^+$ abundance and the green solid line represents gas density. The timescales until the density reaches 10$^5$~cm$^{-3}$ are 1.14$\times$10$^6$~yrs in the cyanopolyyne peak model and 2.14$\times$10$^6$~yrs in the ammonia peak model; the maximum timescales in the horizontal axis are different in two figures.} \label{fig9}
\end{figure*}

We adopt the chemical network used in \citet{Lee04} and the updated N$_2$ binding energy used in \citet{Chen09}.
In the chemical network, the interactions between gas and dust grains (i.e., freeze-out and evaporation of molecules on and off grain surfaces) are included as well as the pure gas chemistry.
We assume that the grain surface is coated by water ice and use the same initial elemental abundances as presented in Table 3 of \citet{Lee04}.
According to previous studies \citep{Hirota98, Hirota03}, the two peaks have a current density of $\sim$10$^5$~cm$^{-3}$.
We therefore consider that the densities of the two peaks evolve from 
10$^3$~cm$^{-3}$ to 10$^5$~cm$^{-3}$ and assume that a core is formed 
when a density condensation is larger than 10$^3$~cm$^{-3}$, which is
the lowest density at which prestellar cores can be identified \citep{Ward-Thompson07}. 
Table~\ref{tbl-12} lists the timescale of density evolution, i.e., the amount of time spent from each density to the next density until the density of the two peaks reaches 10$^5$~cm$^{-3}$ in the models, which fit the observations the best.
The densities in the two models gradually increase as time goes by (see Fig.~\ref{fig9}). 
The timescale spent at low densities (until the density reaches 3$\times$10$^4$~cm$^{-3}$) is shorter in the cyanopolyyne peak model (1.1$\times$10$^6$~yrs) than in the ammonia peak model (2.1$\times$10$^6$~yrs); the density evolves faster in the cyanopolyyne peak model than the ammonia peak model until the density reaches 3$\times$10$^4$~cm$^{-3}$.

\tabletypesize{\small}
\begin{deluxetable}{ccc}[t!]
\tablecolumns{3}
\tablewidth{0pt}
\tablecaption{Timescale of Density Evolution for Chemical Modeling.}\label{tbl-12}
\tablehead{
\colhead{} & \multicolumn{2}{c}{Timescale\tablenotemark{a} (yrs)} \\
\cline{2-3}
\colhead{Gas Density\tablenotemark{b} (cm$^{-3}$)} & \colhead{Cyanopolyyne Peak Model} & \colhead{Ammonia Peak Model}}
\startdata
10$^{3}$          & 1.0$\times$10$^{6}$ & 1.5$\times$10$^{6}$ \\
10$^{4}$          & 1.0$\times$10$^{5}$ & 6.0$\times$10$^{5}$ \\
3$\times$10$^{4}$ & 4.0$\times$10$^{4}$ & 4.0$\times$10$^{4}$ \\
\enddata
\tablenotetext{a}{Timescale spent from a given density to the next density. The final density is 10$^5$~cm$^{-3}$.}
\tablenotetext{b}{Gas density with the mean molecular weight of $\mu=2.3$.}
\end{deluxetable}

Figure~\ref{fig9} presents the abundance evolution of CO and N$_2$H$^+$ in the best-fit models for the cyanopolyyne peak and the ammonia peak, and the abundances of the two molecules at $n(\rm H_2)$=10$^5$~cm$^{-3}$ in the two models are listed in Table~\ref{tbl-13}. 
During the density evolution, the CO freeze-out occurs in the two models, but it is more significant in the ammonia peak model than the cyanopolyyne peak model, as seen in the observations (see Table~\ref{tbl-11}). The abundance of CO in the cyanopolyyne peak model is higher than
that of CO in the ammonia peak model by a factor of $\sim$3 at $n(\rm H_2)$=10$^5$~cm$^{-3}$.
On the other hand, the N$_2$H$^+$ abundances in the two models are similar at $n(\rm H_2)$=10$^5$~cm$^{-3}$; the N$_2$H$^+$ abundance decreases quickly at $n(\rm H_2)$=10$^4$~cm$^{-3}$ in the cyanopolyyne peak model because of its destruction via abundant CO, while the N$_2$H$^+$ abundance declines slowly in the ammonia peak model because its destroyer, CO, is already depleted at the low density. 
In the cyanopolyyne peak model, the maximum abundance of N$_2$H$^+$ is lower than that in the ammonia peak model because its destruction by CO is more efficient in the cyanopolyyne peak model.

According to our modeling, the timescale difference of 1$\times$10$^6$~yrs in density evolution is needed in order to explain the chemical differentiation between the cyanopolyyne and ammonia peaks; the cyanopolyyne peak has a shorter timescale than the ammonia peak. 
This implies that the core at the ammonia peak forms earlier than the core at the cyanopolyyne peak, which is consistent with the suggestion by previous studies; core formation propagated from the ammonia peak to the cyanopolyyne peak \citep{Hanawa94, Howe96, Markwick00, Markwick01, Saito02}. 
However, the current densities of the two peaks are similar. Therefore, once the core at the ammonia peak forms, it might evolve very slowly until the core at the cyanopolyyne peak forms and catches up in its density evolution.
According to our chemical models combined with the density evolution of cores, the more evolved chemistry at the ammonia peak can be attributed to the longer timescale of density evolution at the peak.

\clearpage

\begin{deluxetable*}{ccc}[t!]
\tabletypesize{\scriptsize}
\tablecolumns{3}
\tablewidth{0pt}
\tablecaption{Abundances of CO and N$_2$H$^+$ at $n(\rm H_2)=10^5$~cm$^{-3}$ in the Chemical Calculation.}\label{tbl-13}
\tablehead{
\colhead{} & \multicolumn{2}{c}{Abundance} \\
\cline{2-3}
\colhead{Molecule} & \colhead{Cyanopolyyne Peak Model} & \colhead{Ammonia Peak Model}}
\startdata
CO\tablenotemark{a}          & 6.50$\times$10$^{-5}$   & 2.17$\times$10$^{-5}$ \\
N$_2$H$^+$\tablenotemark{b}  & 1.06$\times$10$^{-10}$   & 1.03$\times$10$^{-10}$ \\
\enddata
\tablenotetext{a}{The maximum CO abundance is 1.2$\times$10$^{-4}$ in both models.}
\tablenotetext{b}{The maximum abundance of N$_2$H$^+$ is different in the two models (5$\times$10$^{-10}$ in the cyanopolyyne peak model and 6.5$\times$10$^{-10}$ in the ammonia peak model).}
\end{deluxetable*}

\citet{Lee03} suggested that the chemical evolution of a core depends on its dynamical timescale as well as its density structure by comparing the chemical distributions in three isolated prestellar cores (L1512, L1544, and L1689B).  According to \citet{Lee03}, L1544 and L1689B have the same density distribution, but L1689B is chemically much younger than L1544. In conclusion, L1689B might experience a free-fall like dynamical process, while L1544
might undergo an ambipolar diffusion process for the dynamical evolution. 
In the TMC-1 filament, the core associated with the ammonia peak requires a longer evolutionary timescale, so it would have formed via  ambipolar diffusion. 
On the other hand, the core associated with the cyanopolyyne peak might have formed by a more violent dynamical process. Some evidence for the dynamical process on a short timescale is present in our observations. First, the 1.2~mm dust continuum emission, which traces the total amount of material the best, shows  a complex structure around the cyanopolyyne peak (see Fig.~\ref{fig1}). Second, the central velocity of the N$_2$H$^+$ line changes sharply around the cyanopolyyne peak (see Fig.~\ref{fig8}). Finally, the molecular line profiles, especially the isolated component of N$_2$H$^+$ $J$=1--0, show a double peak feature (see Fig.~\ref{fig4}), indicative of two velocity components.
(CS 2--1 and C$^{18}$O 2--1 are also affected by optical depths.)
From this evidence, we could consider that two molecular clouds are colliding around the cyanopolyyne peak to induce a fast core formation.

\section{Summary}
\label{sect-6}
We study a possible interpretation of the differentiated chemical distribution in the TMC-1 region using various observational data and chemical modeling.

From $Spitzer$ infrared observations,
we confirm that there is no YSO candidates along the TMC-1 filament
although there are 6 YSO candidates located outside or at boundary of the TMC-1 filament; five classified
YSO candidates objects are classified as Class II sources,
and IRAS source (IRAS 04381+2540) is a Class I object, which agree with previous classifications \citep{Dunham08, Gutermuth08, Gutermuth09, Rebull10}.

The CS 2--1 and N$_2$H$^+$ 1--0 lines are stronger at the ammonia peak than at the cyanopolyyne peak.
On the other hand, C$^{17}$O 2--1 and C$^{18}$O 2--1 lines are stronger at the cyanopolyyne peak than at the ammonia peak.

Using the C$^{17}$O and N$_2$H$^+$ lines, which have hyperfine structures,
we calculate the H$_2$ column density
to compare with the column density calculated from the dust continuum emission,
which traces the material the best along the line of sight.
The column densities calculated from the dust continuum emission
and from the N$_2$H$^+$ 1--0 line are similar at the cyanopolyyne and ammonia peaks, while the column density calculated from C$^{17}$O 2--1 is much greater at the cyanopolyyne peak than at the ammonia peak.

Comparing the depletion factors of CO and  N$_2$H$^+$ (see Table~\ref{tbl-11}), 
we find that the CO molecule is much less depleted at the cyanopolyyne peak than at the ammonia peak, while the N$_2$H$^+$ abundance is similar at both peaks, 
ruling out the possibility of a late-time secondary abundance peak of carbon-chain molecules \citep{Ruffle97, Ruffle99, Li02, Lee03} because both CO and N$_2$H$^+$ are significantly depleted in the late stage of chemical evolution.
Therefore, the relative depletion factors of CO and N$_2$H$^+$ imply that the cyanopolyyne peak is chemically younger (see Section~\ref{sect-1}).

According to our chemical modeling, the ammonia peak stays for a longer time
at low density (10$^3$--10$^5$~cm$^{-3}$) than the cyanopolyyne peak
to explain the chemical difference between the two peaks, indicating that the dynamical evolution can affect
the chemical evolution, resulting in different chemical states even at the same final density.
As a result, we suggest that the chemical differentiation along the TMC-1 filament can be induced
by different dynamical processes of core formation; the ambipolar diffusion for the ammonia peak and a free-fall like process for the cyanopolyyne peak. Some evidence, such as the complex distribution of the 1.2~mm continuum emission (see Fig.~\ref{fig1}), the sharp shift of the central velocity of N$_2$H$^+$ 1--0 (see Fig.~\ref{fig8}), and the double peak feature of molecular line profiles around the cyanopolyyne peak (see Fig.~\ref{fig4}) suggest that two molecular clouds are colliding around the cyanopolyyne peak to induce a fast core formation.

\acknowledgments
This research was supported by the Basic Science Research Program through the National Research Foundation of Korea (NRF) (grant No. NRF-2015R1A2A2A01004769) and the Korea Astronomy and Space Science Institute under the R\&D program (Project No. 2015-1-320-18) supervised by the Ministry of Science, ICT and Future Planning.


\end{document}